\documentclass[allclo]{nonameart}
\usepackage{amsfonts}
\usepackage{amssymb}
\usepackage{amsmath}

\DeclareMathOperator{\rmsign}{sign}

\title{Three-Body Dynamics in One Dimension:\\
A Test Model for the 3-Nucleon System with Irreducible Pionic Diagrams}

\author{T. Melde \instnr{1}, L. Canton\instnr{1,2}, J.P. Svenne \instnr{3}}
\instlist{
Dipartimento di Fisica, Universit\`a di Padova,
Via F. Marzolo 8, Padova, Italy, I-35131
\and Istituto Nazionale di Fisica Nucleare, Sezione di Padova,
Via F. Marzolo 8, Padova, Italy, I-35131
\and
Winnipeg Institute for Theoretical Physics and
Department of Physics and Astronomy,
University of Manitoba,
Winnipeg, MB, Canada R3T 2N2
}
\runningauthor{T. Melde, L. Canton, J.P. Svenne}
\runningtitle{Three-Body Dynamics in One Dimension: A Test Model}
\usepackage{graphicx}
\begin{document}
\maketitle
\begin{abstract}
We formulate the three-body problem in one dimension
in terms of the (Faddeev-type) integral equation approach.
As an application, we develop a spinless, one-dimensional (1-D) model
that mimics three-nucleon dynamics in one dimension.
Using simple two-body potentials that reproduce the deuteron
binding, we obtain that the three-body system binds at about 7.5 MeV.
We then consider two types of residual pionic corrections in the dynamical
equation; one related to the 2$\pi$-exchange three-body diagram, the other
to the 1$\pi$-exchange three-body diagram. We find that the first
contribution
can produce an additional binding effect of about 0.9 MeV. The second
term produces smaller binding effects, which are, however, dependent on
the  uncertainty in the off-shell extrapolation of the two-body t-matrix.
This presents interesting analogies with what occurs in three dimensions.
The paper also discusses the general three-particle quantum scattering
problem, for motion restricted to the full line.
\end{abstract}
\section{Introduction}
The quantum mechanics of one-dimensional systems has always attracted
interest in various research fields and has been extensively studied and
investigated for a variety of purposes, ranging from the study of
exactly solvable many-body systems~\cite{many}, to the phenomenological
description of tunneling phenomena~\cite{tunnel} in quantum wires,
semiconductors, or across junctions~\cite{junction}. The mathematical
properties for the underlying Schr\"odinger operators have been recently
reviewed in  ref.~\cite{Simon2000}.  Quantum scattering in one dimension
has been also considered because of its value in pedagogical
works~\cite{Lip73,GP90,BLA00}, since it retains sufficient complexity to
embody many of the physical concepts and behaviours that occur in the
more complex three-dimensional processes, but without the many technical
aspects required for dealing with quantum scattering in higher dimensions.

One-dimensional models are furthermore often constructed to gain deeper
insight into the methods and approximations introduced to reduce the
complexity of the higher-dimensional cases. In this spirit, we developed
a one dimensional three-body model with the scope of testing
three-nucleon theories in a simplified environment. We are in particular
interested in the formulation of the quantum three-body theory on the
full,  $(-\infty,+\infty)$ line, based on a system of coupled integral
equations for the three-body transition operators  (AGS
equations~\cite{AGS}, extended to an arbitrary number of
  bodies by Grassberger
and Sandhas~\cite{GS67}). The approach is directly linked to the first
mathematical theory of three-body scattering by Faddeev~\cite{Fadd}, based
on the integral formulation for the three-body Green's function and
extended to N bodies by Yakubovski~\cite{Yak67}.

The formulation of the one-dimensional three-body problem in terms of
the Faddeev-AGS scattering theory represents a somewhat ``unusual''
problem, since one-dimensional scattering corresponds also to the
potential-barrier problem, which therefore has to be implemented with
the clustering structure of the three-body problem.   We illustrate in
this paper an approach that harmonizes the typical two-channel structure
of the ``tunneling'' problems, due to the presence of the transmission
and reflection processes, with the cluster structure of the three-body
problem, thus leading to an overall reduced S-matrix
for the two-fragment processes which is
$6\times 6$. The formulation presented herein, with straightforward
modifications, is suited also to describe the tunneling (or
potential-barrier) effects due to the scattering of a two-body
one-dimensional cluster impinging on an external barrier potential.
A 3-D application along this line already has been discussed for a
molecular system~\cite{HK99}.

The present work was motivated by the need to develop a simplified, 1-D
test model for the non-relativistic treatment of the system consisting of
  three
nucleons interacting via exchanges of a force-mediating ``pion''  (plus
some shorter-range contributions). In modern three-nucleon dynamics,
one of the crucial problems involves the treatment of those additional
mesonic contributions that cannot be effectively described by the
standard nucleon-nucleon  potentials. Such terms are generally referred
to as irreducible three-nucleon-force (3NF) contributions and are used
to construct a phenomenological 3N potential.  The hope is to cure in
this way the discrepancies observed between experiments and rigorous
Faddeev three-body calculations with realistic 2N
potentials~\cite{Paris,Bonn}.  The subsequent employment of the newer
high-precision 2N potentials~\cite{Nij94,Arg95,CDBonn} essentially
confirmed the existence of such discrepancies. For example, the
three-nucleon bound state, the triton, is typically under-bound by the
high-precision NN potentials and the vector analyzing powers are
substantially underestimated for low-energy nucleon-deuteron scattering
(the so-called $A_{y}$-puzzle)~\cite{PhysRept,Knu98}.

Considerable progress has been made since the introduction of 3NF based
on isobars by Fujita and Miyazawa~\cite{FM57}, leading to a number of
different 3NFs~\cite{TMF,Brasil,OrK92,van94} in use today, that can all
be adjusted to describe adequately the triton binding energy.
However, none of these terms have solved also the  $A_{y}$-puzzle, which
suggests that work still needs to be done to understand the role of
the 3NF terms in few-nucleon systems. A few years ago, an approach has
been developed~\cite{Can98} that reduces the field theoretic problem of
the pion-three-nucleon dynamics  to an approximate set of integral
equations for the coupled $\pi$NNN-NNN system. The method extends the
standard AGS formulation to the case of a three-body system with the
additional pionic degree of freedom. The resulting equations merge
together the Faddeev structure for the three-nucleon sector with the
Yakubovski structure of the four-body sector, and has been shown to
possess a kernel which is connected after iterations.

Recently~\cite{CMS01}, a practical approximation scheme for the
treatment of this coupled $\pi$NNN-NNN system has been proposed. The
scheme is based on the assumption that the dominant part of the
pion-exchange processes can be effectively described by contributions to
the 2N potential, while the residual mesonic aspects lead to the
addition of smaller, correction-type, 3NF effects. This assumption forms
the basis for the standard description of nuclear systems in terms of
aggregates of (pairwise) interacting nucleons, and ultimately relies on
the chiral nature of the hadron interactions~\cite{van94}.  The
irreducible 3NF corrections due to the residual pion dynamics are
treated perturbatively in the extended AGS equation, which is further
reduced to an effective two-body integral equation in the inter-cluster
momentum variable, by means of a finite-rank expansion of the two-nucleon
potential.

At the lowest order, the residual pion dynamics produced irreducible
correction contributions to the driving term of the AGS equation, and
one type of these corrections can be identified with the so-called
$2\pi$-exchange 3NF diagram, and leads to the force models discussed,
e.g., in  refs.~\cite{FM57,TMF,Brasil,OrK92,van94}. But at the same time, the
approach produced also another type of irreducible pionic
diagrams~\cite{CMS01,CS00} --~the $1\pi$-exchange 3NF term~-- that could not be
identified with any previously known 3NF expression and which, as shown in
completely realistic situations~\cite{CS01a,CSH02}, has the potential to
solve the $A_y$ problem. For a brief review, see also ref.~\cite{CPS01}.

In the following, we develop a one-dimensional spinless model that
mimics three-nucleon dynamics in one dimension.
In particular, we investigate numerically the effects of these
irreducible  ``pionic'' corrections in the one-dimensional three-body
bound state by solving the homogeneous version of the AGS equation.
This 1-D model was originally constructed as a laboratory to investigate
the dynamical effects of the meson on the three-body bound
state~\cite{Mel01}. However, one-dimensional systems in themselves are of
considerable interest~\cite{Miy00,Ryb01,AKM01} and in many ways
different from the three-dimensional case. This difference manifests
itself, for example, in the existence of only two partial
waves~\cite{BLA00,NR96,BLA01} and the different structure of Levinson's
theorem~\cite{SB94,BGK85,BGW85}.

Section~\ref{sec:AGS} presents the AGS form of the three-body theory  on the
line, and its reduction to a Lovelace-type equation for a  symmetrized
three-particle system interacting via separable potentials.
In Sect.~\ref{sec:corr} we define the explicit form of
the pionic correction terms to
the  Lovelace-AGS equation. In Sect.~\ref{sec:homo} we consider the
homogeneous version of the three-body integral equation, which leads to
the calculation of the binding energies.
In Sect.~\ref{sec:num} we discuss the details of the three-body model
and illustrate the results obtained. For the conclusions and a summary
of the results we refer to Sect.~\ref{sec:sum}.

Finally, we have also endowed this work with three appendices, devoted
to specific aspects of the one-dimensional two-body system.
Appendix~A illustrates the integral formulation of the
one-dimensional two-body scattering problem. Although the treatment can
be found in the literature, e.g. in ref.~\cite{BLA00},  the discussion
presented in the Appendix stresses the analogies  with the three-body
problem. Also the low-energy expansion of  the two-body amplitudes is
discussed therein.
Appendix~B illustrates the details for
the expansion of the 2-body potentials in a finite-rank form, along the
lines of the Unitary Pole Expansion (UPE) method.
Appendix~C
discusses the method we have employed to solve numerically the
one-dimensional two-body problem in one dimension, for both bound-state
and scattering regimes.
\section{The Alt-Grassberger-Sandhas Three-Body Theory on the Full Line}
\label{sec:AGS}
\subsection{Distinguishable Particles}
In this section we formulate the Faddeev-AGS theory for three
distinguishable  particles moving in one dimension, $(-\infty,+\infty)$.
We consider three spinless particles interacting only  with 2-body
potentials, and assume that these potentials are short-range, or at most
exponentially decaying when the inter-particle  distance goes to
infinity. The case of irreducible 3NF-type diagrams will  be discussed
in Sect.~\ref{sec:corr}.
For simplicity, the two-body potential is assumed
to support only one bound-state, as happens for the 2-nucleon system,
but the formulation can be easily generalized to the case of a finite
number  of bound states. The particles are labeled ``1'', ``2'', and
``3'', and the indices $a,b,c,\dots$ range over the three labels. We
will adhere to the  odd-man-out notation which implies that when $a=1$,
for example,  $V_a$ denotes the interaction between the pair $(23)$.

A system of three particles is described by the Hamiltonian
\begin{equation}
H=H_0+\sum_a{
V_a} \, .
\end{equation}
We then introduce the channel states, which are eigenstates of the channel
Hamiltonians
\begin{equation}
H_a=H_0+V_a \, ,
\end{equation}
with $a=1,2,3$. The full
Hamiltonian can also be written as
\begin{equation}
H=H_a+V^a \, ,
\end{equation}
where
\begin{equation}
V^a=\sum_{b\ne a}
{
V_b
}
\end{equation}
denotes the interaction ``external'' to channel $a$. In the one-dimensional
system the channel states can be described by the following wave functions
\begin{equation}
\Phi_{\pm a}=\Phi_{\pm q_a}\left(x_a,y_a\right)
=e^{\pm \imath q_ay_a}\phi_a\left(x_a\right) \, ,
\end{equation}
where $\phi_a\left(x_a\right)$ denotes the properly normalized bound-state
wave function of the two-particle fragment and
$e^{\pm \imath q_ay_a}$ denotes
the relative motion of the particle $a$ with respect to the c.m.
of the two-particle fragment $\left(bc\right)$.
Here, $x_a,y_a$ are the Jacobi coordinates
for partition $a\left(bc\right)$ and $q_a$ is the relative momentum between
particle $a$ and the cluster $\left(bc\right)$. The $\pm$ in the channel
wave function is due to the presence of two
physically different states, one with an incoming wave from the
left and one with an incoming wave from the right. Note that this definition
implicitly assumes $q_a\ge 0$. Alternatively, we define also
\begin{equation}
\Phi_{a}=\Phi_{q_a}\left(x_a,y_a\right)
=e^{\imath q_ay_a}\phi_a\left(x_a\right)\, ,
\end{equation}
where $q_a$ may assume also negative values.

The channel wave functions are eigenstates of the channel Hamiltonian $H_a$
with the eigenvalue
\begin{equation}
E_a=\frac{\hbar^2{q_a}^2}{2M_a}+\hat E_a \, ,
\end{equation}
where $\hat E_a$ is the bound state energy of the two-particle fragment
and $M_a$ is the reduced mass of particle $a$ and the
cluster $\left(bc\right)$. The masses of the three (identical) particles
are given by $m_N c^2=938.27$~MeV, the mass of the pion is
$m_\pi c^2=134.98$~MeV and the reduced mass of the particles
$b,c$ is $\mu_a$. Furthermore,
because we consider three identical particles we also have
$\mu_a=\mu_b=\mu_c=\mu$ and $M_a=M_b=M_c=M$.

Let us assume an asymptotic 1-D plane wave incoming from the left,
described by the channel wave functions $\Phi_{+a}$. Then we can have three
different transmitted two-cluster waves, and similarly for the reflected
terms, which leads to the asymptotic form
\begin{equation}
\Psi_{a}^L\left(x,y\right)\rightarrow
\begin{cases}
\Phi_{+b}\delta_{ab}+R^L_{ab}\left( E\right)\Phi_{-b},
& \text{for $  y_b\rightarrow -\infty$}\\
T^L_{ab}\left( E\right)\Phi_{+b},
& \text{for $ y_b\rightarrow +\infty$}
\end{cases}
\end{equation}
where $x,y$ is a generic set of Jacobi coordinates.
The superscript $L$ reflects the fact that the incident wave is coming from
the left. Differently from the two-body case
where there is only one single channel, we are now in a
multichannel situation where the rearrangement processes
(i.e., $a\ne b$) have
to be also described. This leads  to the definition of the reduced
S-matrix for two-fragment scattering which is formally equivalent to the
one defined in
Appendix~A for two-particle scattering,
but features in addition the cluster labels,
\begin{equation}
{\bf \tilde S_{ab}}\left(E\right)=
\begin{pmatrix}
T^L_{ab}\left(E\right) & R^R_{ab}\left(E\right) \\
R^L_{ab}\left(E\right) & T^R_{ab}\left(E\right)
\end{pmatrix} \, .
\label{eq:TR3space}
\end{equation}
This definition produces a $6\times 6$ scattering matrix connecting all
six possible ``two-fragment asymptotic channels'' $\Phi_{\pm a}$. The
occurrence of such type of S-matrix has been observed already in the
framework of an exactly soluble problem in ref.~\cite{McG64}

To solve the one-dimensional three-body scattering problem
the S-matrix has to be determined, and for this we use
the known expression~\cite{Taylor}
\begin{equation}
\left\langle \Phi_a\right|S_{ab}\left|\Phi_b\right\rangle=
\delta_{ab}\delta\left(q_a-q_b\right)-2\pi i \delta\left(E_a-E_b\right)
\left\langle \phi_a\right| V^a \left| \Psi_b \right\rangle
\, .
\label{eq:STaylor}
\end{equation}
It is then possible to introduce~\cite{Gloeckle} the transition operators
$U_{ab}$, with the definition
\begin{equation}
\left\langle \Phi_a\right|U_{ab}\left| \Phi_b\right\rangle=
\left\langle \Phi_a\right|V^a\left| \Psi_b\right\rangle \, ,
\end{equation}
which satisfy the operatorial Alt-Grassberger-Sandhas equation
\begin{equation}
U_{ab}=\bar \delta_{ab}G_0^{-1}+\sum_{c}{
\bar \delta_{ac}V_c
G_c U_{cb} \,
}
\, ,
\label{eq:AGS}
\end{equation}
where $\bar \delta_{ab}=1-\delta_{ab}$ and
$G_a\left(z\right)=\left(z-H_a\right)^{-1}$.
To recover the $6\times 6$ structure for the reduced S-matrix
we decompose the momentum on the line
\begin{equation}
q=\left| q\right|\rmsign\left(q\right) \, ,
\end{equation}
where $\rmsign\left(q\right)$ denotes the sign of $q$. Then, we consider
the two delta functions in the expression of the S-matrix,
Eq. (\ref{eq:STaylor}),
\begin{equation}
\delta\left(q_a-q_b\right)=\delta\left(\left|q_a\right|-
\left|q_b\right|\right)
\delta_{\rmsign\left(q_a\right)\rmsign\left(q_b\right)} \, ,
\end{equation}
and
\begin{equation}
\delta\left(E_a-E_b\right)=\frac{2M}{\hbar^2}
\delta\left(\left|q_a\right|^2-\left|q_b\right|^2\right)
=\frac{M}{\hbar^2\left|q_b\right|}
\delta\left(\left|q_a\right|-\left|q_b\right|\right) \, .
\end{equation}
In this last equation we have assumed for simplicity
that the bound-state energies of any two-particle fragments are the same.
The generalization to the case with different binding energies,
{i.e.} $\hat E_a \ne \hat E_b$ for $a\ne b$, is straightforward.
Therefore, the S-matrix can be written as
\begin{equation}
\left\langle \Phi_a\right|S_{ab}\left|\Phi_b\right\rangle=
\delta\left(\left|q_a\right|-\left|q_b\right|\right)
\left\langle \Phi_{\pm a}\right|\tilde S_{ab}\left|\Phi_{\pm b}\right\rangle
\, ,
\end{equation}
where
\begin{multline}
\label{bigS}
  {\bf \tilde S_{ab}}\left(E\right)  =
\left\langle \phi_{\pm a}\right|\tilde S_{ab}\left|\phi_{\pm b}\right\rangle
  \\
  =
\delta_{ab}
\begin{pmatrix}
1 & 0\\
0 & 1\\
\end{pmatrix}
-2\pi \frac{\imath M}{\hbar^2\left|q_b\right|}
\begin{pmatrix}
t_{ab}\left(+\left|q_a\right|,+\left|q_b\right|;E\right) &
t_{ab}\left(+\left|q_a\right|,-\left|q_b\right|;E\right) \\
t_{ab}\left(-\left|q_a\right|,+\left|q_b\right|;E\right) &
t_{ab}\left(-\left|q_a\right|,-\left|q_b\right|;E\right)
\end{pmatrix} \, .
\end{multline}
To solve the AGS equation we expand the 2-body potentials
into a finite-rank form. The method of expansion
is discussed in
Appendix~B.
For simplicity, in this section we illustrate the formulation for
the case of a separable (rank-one) expression for the 2-body potentials
\begin{equation}
V_a^{sep}=\frac{V_a|\phi_a\rangle \langle\phi_a|V_a}{
\langle \phi_a | V_a | \phi_a \rangle}
\, ,
\end{equation}
where $|\phi_a\rangle$ corresponds to the properly normalized bound-state
of the two-body subsystem $a$. Then the AGS equation transforms into
an effective two-cluster Lovelace-type equation
for the transition amplitudes \cite{Lov64},
which is an integral equation in the inter-cluster momentum variable,
\begin{multline}
X_{ab}\left(q_a,q_b;E\right) = Z_{ab}\left(q_a,q_b;E\right)
\\
+\sum_{c}{
\int_{-\infty}^{+\infty}{
Z_{ac}\left(q_a,q_c;E\right)\Delta_c \left(q_c;E\right)
X_{cb}\left(q_c,q_b;E\right)
dq_c}
} \, .
\end{multline}
\begin{figure}[hpt]
\centerline
{
\includegraphics[width=7cm]{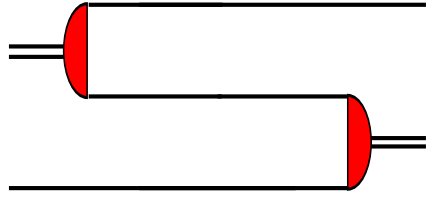}
}
\caption{The exchange diagram, representing the AGS driving term}
\label{fig:FeynAGS}
\end{figure}
The driving term of this equation corresponds to the exchange diagram
represented in Fig.~\ref{fig:FeynAGS},
and is calculated according to the expression \cite{SZ74}
\begin{equation}
Z_{ab}\left( {q_a,q_b,E} \right)=
\frac{
\chi_a\left(\frac{q_a}{2}+q_b \right)
\chi_b\left(-q_a-\frac{q_b}{2} \right)
}{
E-\frac{\hbar ^2}{2m_N}\left(
q_a^2+\left(q_a+q_b \right)^2+\left(q_b \right)^2\right)}
     \label{eq:ZAGST} \, .
\end{equation}
Note that there is also a second driving term $Z_{ac}$ with
$a\ne c\ne b\ne a$ and implying
the appropriate Jacobi-coordinate transformations.
The $\Delta_a \left(q_a;E\right)$ is given by
\begin{equation}
\Delta_a \left(q_a;E\right) =  \left[{
\int_{-\infty}^{+\infty} \phi_a(p) \chi_a(p) dp
-
\int_{-\infty}^{+\infty} {\chi_a(p) \chi_a(p) dp \over
{E-{{\left( {\hbar q_a} \right)^2} \over {2M}}-{{\left( {\hbar p} \right)^2}
\over {2\mu }}}}}\right]^{-1} \, .
     \label{eq:DAGS}
\end{equation}
In these last equations we have also introduced the form factors of
the separable interaction,
$|\chi_a\rangle = V_a |\phi_a\rangle$, for convenience.

Once evaluated for the on-shell momenta, the solution of
this integral equation provides the transition amplitudes
\begin{equation}
\left\langle \phi_{\pm a}\right|U_{ab}\left| \phi_{\pm b}\right\rangle  =
X_{ab}\left(\pm |q_a|,\pm |q_b|; E\right)=t_{ab}(\pm |q_a|,\pm |q_b|; E)
  \, ,
\end{equation}
which are needed in Eq.~(\ref{bigS}) for the construction of the
S-matrix.
\subsection{Symmetrization}
Lovelace~\cite{Lov64} has discussed how the system of coupled equations can
be symmetrized, leading to a single-channel integral equation
\begin{equation}
{\cal X} \left(q,q';E\right)={\cal Z} \left(q,q';E\right)
  +\int_{-\infty}^{+\infty}
  {{\cal Z} \left(q,q'';E\right) \Delta \left(q'';E\right)
{\cal X} \left(q'',q';E\right)dq''}\, .
\label{eq:Lovelace}
\end{equation}
The $q,q'$, and $q''$ denote the inter-cluster momentum  variables of
the two-fragment states. The function $\Delta
\left(q'';E\right)$ is the same as calculated  in Eq.~(\ref{eq:DAGS}),
while the
symmetrized driving term  ${{\cal Z} \left(q,q'';E\right)}$ is now twice
that calculated in Eq.~(\ref{eq:ZAGST}). The method works for both bosons
and fermions, the difference between the two being in the representation
of the  two-body sub-amplitude. For bosons, the two-body t-matrix has to  be
expanded via the UPE method using only symmetric states, i.e., with
the even functions $\chi_i(p)=\chi_i(-p)$.
In the case of three identical particles only one type of two-cluster
partition is observable, and therefore the $6\times 6$ structure of the
previously defined S-matrix and of the
transition amplitude collapses back into a $2\times 2$
structure, as was for the case of
two-body scattering.

The ${\cal X} \left(q,q';E\right)$ represent the  two-fragment
transition amplitude which has to be evaluated, for a given energy $E$,
at the only two possible on-shell values for $q,q'$.
These momenta have to satisfy the condition
\begin{equation}
q^2=\left.q'\right.^2=\frac{2M }{\hbar^2}\left( E-E_D\right)=
\frac{2M}{\hbar^2}\bar E \, ,
\end{equation}
where $E_D$ is the bound state energy of
the two-particle fragment.
Because the system is defined on the real line, the values for the
inter-cluster momenta have to be
\begin{equation}\left|q\right|=\sqrt{\frac{2M}{\hbar^2}\left(E-E_D\right)}
=\sqrt{\frac{2M \bar E}{\hbar^2}}
\, .
\end{equation}
This allows us to write the S-matrix for one-dimensional $2+1$ scattering
in the following way
\begin{equation}
S\left( E\right)=
\begin{pmatrix}
1 & 0\\
0 & 1\\
\end{pmatrix}
-2\pi\frac{\imath M}{\hbar^2 \left|q\right|}
\begin{pmatrix}
{\cal X} \left( \left|q\right|,\left|q\right|; E\right)&
  {\cal X}  \left(\left|q\right|,-\left|q\right|; E\right)\\
{\cal X}  \left(-\left|q\right|,\left|q\right|; E\right)&
  {\cal X}  \left(-\left|q\right|,-\left|q\right|; E\right)\\
\end{pmatrix}
\, ,
\end{equation}
which has the same structure as the two-body S-matrix defined in
Appendix~A.

\section{Treatment of the Irreducible Pionic Corrections}
\label{sec:corr}
The method, discussed up to now for a separable (i.e., rank one) expression
for the two-body interaction, has to be extended to a rank-$N$ expansion,
as described in
Appendix~B.
The symmetrized Lovelace-type equation for a rank-$N$ potential is given by
\begin{multline}
{\cal X}_{ij} \left(q,q';E\right)={\cal Z}_{ij} \left(q,q';E\right)
\\
  +\sum_{k,l}^N \int_{-\infty}^{+\infty}
  {{\cal Z}_{ik} \left(q,q'';E\right) \Delta_{kl} \left(q'';E\right)
{\cal X}_{lj} \left(q'',q';E\right)dq''}\, ,
\label{eq:rankN}
\end{multline}
where $\Delta _{kl}$ is defined  in
Appendix~B.
The driving terms acquire a matrix structure in the rank space,
${\cal Z}_{ij}$.

Typically, the symmetrized Lovelace equation leads to a driving term
\begin{equation}
{\cal Z}_{ij}=2Z_{ij}^{N}
\, ,
\end{equation}
which is twice that calculated in Eq.~(\ref{eq:ZAGST}).
The label ``N'' in the above expression is a reminder of the
off-diagonal character of the driving term, due to the
presence of $\bar\delta_{ab}$ in Eq.~(\ref{eq:AGS}). The factor 2
arises because in the symmetrized Lovelace three-body equation there
are two off-diagonal elements that add coherently.

The three-body Lovelace-type integral equation can be extended to
include the degrees of freedom of one single meson~\cite{Can98}. To
reduce the complexity of the resulting equation, one can consider a
simplified formulation~\cite{CMS01} where the effects of the  two-body
interactions are treated exactly, and the residual irreducible
(3NF-like) mesonic effects are treated  approximately, as
``first-order'' corrections.  The approximation scheme leads to mesonic
corrections to the driving term, Eq.~(\ref{eq:ZAGST}),
  of the standard three-body
equation with two-body potentials  discussed in the previous section.

These pionic corrections
corrections modify the driving term according to the expression
\begin{equation}
{\cal Z}_{ij}=Z_{ij}^{D}+2Z_{ij}^{N} \, ,
\end{equation}
where the superscript $D$ denotes now the presence of  diagonal-type
contributions.
Without these corrections one recovers the standard AGS-term by considering
only the off-diagonal elements~\cite{SZ74},
i.e., $Z_{ij}^D=0$, and $Z_{ij}^{N}=Z_{ij}^{AGS}$.

\begin{figure}[hpt]
\centerline
{
\includegraphics[width=7cm]{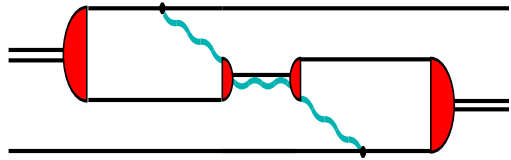}
}
\caption{Correction terms to the off-diagonal AGS driving term}
\label{fig:FeynTM}
\end{figure}

\begin{figure}[hpt]
\centerline
{
\includegraphics[width=7cm]{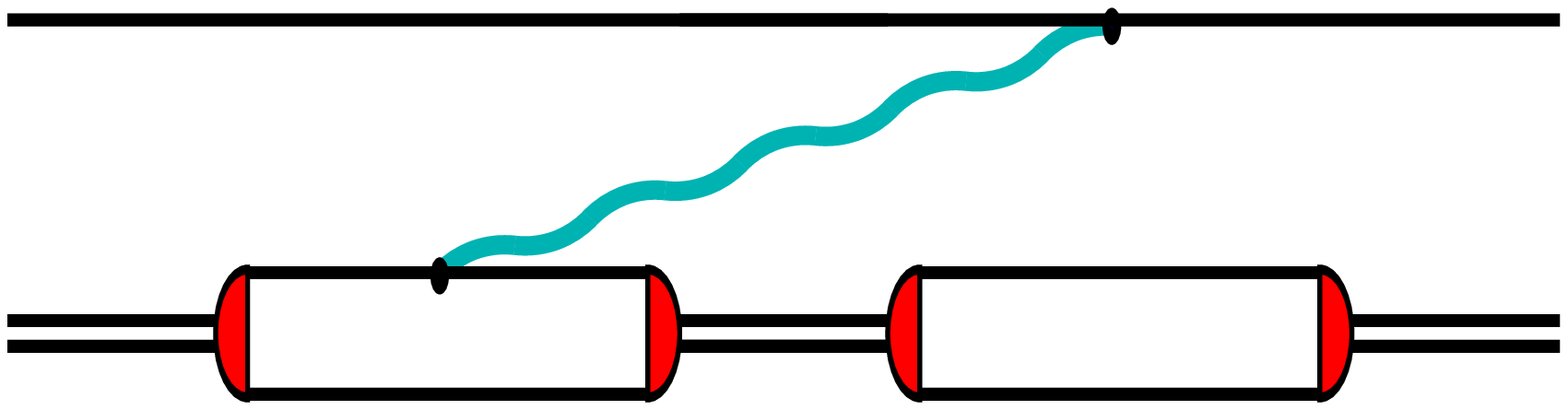}
}
\caption{Correction terms to the diagonal AGS driving term.}
\label{fig:Feyndiag}
\end{figure}

In particular, the standard AGS term in the $N$-rank case
is given by
\begin{equation}
Z_{ij}^{AGS}\left( {q,q',E} \right)={{\chi _i\left( {{q \over 2}+q'} \right)
\chi _j\left( {-q-{{q'} \over 2}} \right)} \over {E-{{\hbar ^2} \over {2m_N}}
\left( {q^2+\left( {q+q'} \right)^2+\left( {q'} \right)^2} \right)}}
\, ,
\label{eq:ZAGS}
\end{equation}
where the form factors $\chi _{i},\chi _{j}$ are defined in
Appendix~B. We remind also that, since we have assumed a
bosonic structure of the particles, only the even parity form factors are
included in the separable expansion.\\
The treatment of the irreducible pionic effects leads one to consider,
in first approximation, the following correction terms
\begin{equation}
Z_{ij}^{N}=Z_{ij}^{AGS}+Z_{ij}^{TPE3}
\end{equation}
\begin{equation}
Z_{ij}^{D}=Z_{ij}^{OPE3} \, .
\end{equation}
The diagrams generating these two additional terms are reported in
Figs.~\ref{fig:FeynTM}~and~\ref{fig:Feyndiag},  for $Z^{TPE3}$  and
$Z^{OPE3}$, respectively (we recall that the standard AGS exchange
diagram was reported in Fig.~\ref{fig:FeynAGS}).

The off-diagonal correction terms are given by
\begin{multline}
Z_{ij}^{TPE3}\left( {q,q',E} \right)=
\\
\int {{{\chi _i\left( p \right)}
\over
{E-{{\left( {\hbar q} \right)^2} \over {2M}}-{{\left( {\hbar p} \right)^2}
\over {2\mu }}}}V^{TPE3}\left( {p,p',q,q'} \right)
{{\chi _j\left( {p'} \right)} \over {E-{{\left( {\hbar q'} \right)^2}
\over {2M}}-{{\left( {\hbar p'} \right)^2} \over {2\mu }}}}dpdp'} \, ,
\end{multline}
and similarly for the diagonal correction terms,
\begin{multline}
Z_{ij}^{OPE3}\left( {q,q',E} \right)=
\\
\int {{{\chi _i\left( p \right)}
\over {E-{{\left( {\hbar q} \right)^2} \over {2M}}-{{\left( {\hbar p}
\right)^2} \over {2\mu }}}}V^{OPE3}\left( {p,p',q,q'}
\right){{\chi _j\left( {p'} \right)} \over {E-{{\left( {\hbar q'} \right)^2}
\over {2M}}-{{\left( {\hbar p'} \right)^2} \over {2\mu }}}}dpdp'} \, .
\end{multline}
The first contribution to the driving term ${\cal Z}_{ij}$ is generated
by an intermediate $\pi$N re-scattering process, as depicted by the diagram in
Fig.~\ref{fig:FeynTM}. In the static approximation,
the amplitude for such a process is given by the expression
\begin{equation}
V^{TPE3}\left( {p,p',q,q'} \right)=
{{f^2} \over \pi }{{\hat t_{\pi N}} \over {\left( {Q^2+\left( {{{m_\pi c}
\over \hbar }} \right)^2} \right)\left( {\left(Q'\right)^2
+\left( {{{m_\pi c}
\over \hbar }} \right)^2} \right)}} \, .
\end{equation}
Here, the $Q,Q'$ are the pion momenta in the center of mass system with the
proper set of Jacobi momenta
\begin{equation}
Q=p-\frac{q}{2}-q'
\end{equation}
\begin{equation}
Q'=-q-p'-\frac{q'}{2}
\end{equation}
and $\hat t_{\pi N}$ denotes the $\pi$N amplitude.

One serious difficulty inherent to this test model was connected with
the construction of this one-dimensional $\pi$N amplitude, since this
represents a completely abstract entity, which has no direct link to the
phenomenological, three-dimensional amplitude of the real  world
\footnote{It should be noted that, in obtaining the 3NFs in the full
three-dimensional form, this $\pi$N scattering amplitude is also the
main source of uncertainty due to the need to find a satisfactory
off-shell extension and suitable pion form factors \cite{Brasil}.}. We
decided to circumvent this difficulty by observing that the $\pi$N
scattering lengths are roughly between one and two orders of magnitude
smaller than the corresponding 2N scattering lengths~\cite{EW}. This
suggests that probably the most simple parametrization for the $\pi$N
amplitude in one dimension is to take the low-energy limit of the
corresponding 2N one-dimensional amplitude, and to divide  this
amplitude by a factor $c_1$ between $10$ and $100$.  The low-energy
expansion for the two-body t-matrix is discussed in
Appendix~A, to which we refer for the details.
The result leads to the following
low-energy structure for the 1-D $\pi$N amplitude
\begin{equation}
\hat t_{\pi N}\left( Q,Q'\right) \approx  -\frac{1}{c_1}\left[
\left(\hat \alpha_0+\hat \alpha_1\right)\left|Q\right|\left|Q'\right|
+\left(\hat \alpha_0-\hat \alpha_1\right)QQ'\right]
\, ,
\label{low-exp}
\end{equation}
where the parameter $c_1$ has been arbitrarily set to $c_1=-15$,
to be approximately consistent with the ratio between the observed
scattering lengths of the $\pi$N and NN systems.

Therefore, the final result for this contribution has the following structure
\begin{eqnarray}
Z_{ij}^{TPE3}\left( {q,q',E} \right) & = &
-\frac{f^2}{c_1\pi}
\left(\hat \alpha_0+\hat \alpha_1\right)
\int {
\frac{\left|Q\right|}{Q^2+\left(\frac{m_{\pi}c}{\hbar}\right)^2}
\frac{\chi_i\left(p\right)}{E-\frac{\left(\hbar q\right)^2}{2M}-
\frac{\left(\hbar p\right)^2}{2\mu}}
dp}
\nonumber
\\
& &
\times
\int {
\frac{\left|Q'\right|}{\left(Q'\right)^2+\left(\frac{m_{\pi}c}{\hbar}\right)^2}
\frac{\chi_j\left(p'\right)}{E-\frac{\left(\hbar q'\right)^2}{2M}-
\frac{\left(\hbar p'\right)^2}{2\mu}}
dp'}
\nonumber
\\
& &
-\frac{f^2}{c_1\pi}
\left(\hat \alpha_0-\hat \alpha_1\right)
\int {
\frac{Q}{Q^2+\left(\frac{m_{\pi}c}{\hbar}\right)^2}
\frac{\chi_i\left(p\right)}{E-\frac{\left(\hbar q\right)^2}{2M}-
\frac{\left(\hbar p\right)^2}{2\mu}}
dp}
\nonumber
\\
& &
\int {
\frac{Q'}{\left(Q'\right)^2+\left(\frac{m_{\pi}c}{\hbar}\right)^2}
\frac{\chi_j\left(p'\right)}{E-\frac{\left(\hbar q'\right)^2}{2M}-
\frac{\left(\hbar p'\right)^2}{2\mu}}
dp'}
\, .
\end{eqnarray}

We stress that the arguments to
arrive at the form Eq.~(\ref{low-exp}) for the subtracted $\pi$N
amplitude have been ultimately quite arbitrary; however they
reproduce the net result that, in the limit $c_1\rightarrow
\infty$, that is when the $\pi$N scattering length is much smaller than
the 2N scattering length, the corresponding 3NF effect vanishes.
The second contribution, $V^{OPE3}$, is generated by the diagram in
Fig.~\ref{fig:Feyndiag}.
The corresponding amplitude is obtained by summing over four terms,
namely~\cite{CS00}
\begin{eqnarray}
  V^{OPE3}\left( p,p',q,q' \right)&=&
  f_1G_0^{\left( 4 \right)}\tilde t_{12}G_0^{\left( 4 \right)}f_3^{\dagger}
+f_2G_0^{\left( 4 \right)}\tilde t_{12}G_0^{\left( 4 \right)}f_3^{\dagger}
\nonumber \\
& &
+f_3G_0^{\left( 4 \right)}\tilde t_{12}G_0^{\left( 4 \right)}f_1^{\dagger}
+f_3G_0^{\left( 4 \right)}\tilde t_{12}G_0^{\left( 4 \right)}f_2^{\dagger}
\, ,
\end{eqnarray}
where $G_0^{\left( 4\right)}$ is the 4-particle free propagator and
$f_{i}$ denotes the creation operator of a pion on nucleon
$i$, while $f_{i}^{\dagger}$ denotes the destruction of a pion on nucleon
$i$. The quantity $\tilde t_{12}$ describes the two-body t-matrix for
nucleons $1,2$. Similarly to the case of the TPE3-term,
also in this case one has to include a cancellation effect, since
one must subtract from the two-body amplitude
the corresponding ``Born terms'',
i.e. the meson-exchange contributions, which can be identified
with the two-body potential itself. Furthermore, the two-body t-matrix
entering in this diagram has to be evaluated in a region which is
substantially off-shell, because of the role of the exchanged meson
in this diagram.
Therefore, following the studies of refs.~\cite{CS00,CS01a,CSH02} we have
considered the effective parameter that governs the cancellation
for the diagonal correction terms:
\begin{equation}
\tilde t_{12}\left( {p,p';z}\right)=c_2
t_{12}\left( {p,p';z} \right)-v_{12}\left( {p,p'} \right) \, ,
\label{eq:cancel}
\end{equation}
where
\begin{equation}
z=E-\frac{\left(\hbar q\right)^2}{2M}
-\hbar c\omega _\pi \left( Q \right)  \, .
\end{equation}
The effective parameter $c_2$
represents an overall correction factor for the far-off-the-energy-shell
extrapolation of the t-matrix \cite{CS01a,CSH02}.
Also, $v_{12}$ denotes the meson exchange contributions as defined in the
nucleon-nucleon potential acting between nucleons $1,2$ and
\begin{equation}
\omega _\pi \left( Q \right)=\sqrt {Q^2+\frac{m_\pi ^2c^2}{\hbar ^2}}
\end{equation}
is the relativistic pion energy with the pion momentum
\begin{equation}
Q=q-q'  \, .
\end{equation}
Using a
static approximation for the four-body Green's functions $G_0^{\left( 4
\right)}$ and inserting the explicit expressions for the creation and
destruction operators gives the following form for the static
contribution to the diagonal correction terms
\begin{multline}
V^{OPE3}\left( {p,p',q,q'} \right)
  ={{f^2} \over \pi }
{2 \over {\hbar c\omega^{3} _\pi \left( Q \right)}}
  \\
\times
\left[ {\tilde t_{12}\left( {p,p';E-{{\left(\hbar q\right)^2} \over {2M }}-
\hbar c\omega _\pi \left( Q \right)} \right)+\tilde t_{12}
\left( {p,p';E-{{\left( {\hbar q'} \right)^2} \over {2M }}-
\hbar c\omega _\pi \left( Q \right)} \right)} \right]
\, .
\end{multline}
In this paper we are using the unitary-pole expansion and consequently the
subtracted NN t-matrix is defined as
\begin{multline}
\tilde t_{12}\left( {p,p';E-{{\left(\hbar q\right)^2}
\over {2M }}-\hbar c\omega _\pi
\left( Q \right)} \right)=
\\
\sum\limits_{ij} {\chi _i\left( p \right)
\left(
c_2
{\Delta _{ij}\left( {E-{{\left(\hbar q\right)^2}
\over {2M }}-\hbar c\omega _\pi
\left( Q \right)} \right)+\eta_{i} ^{-1}\delta _{ij}}
\right)
\chi _j\left( {p'} \right)}
\, ,
\end{multline}
which leads to the final expression for the
diagonal contribution of the
correction term in a rank-$N$ separable expansion
\begin{multline}
Z_{ij}^{OPE3}\left( {q,q',E} \right)=
{{f^2} \over \pi }{2 \over {\hbar c\omega ^{3}_\pi
\left( Q \right)}}\left\{ {\sum\limits_{kl}
{\int {{{\chi _i\left( p \right)\chi _k\left( p \right)}
\over {E-{{\left( {\hbar q} \right)^2} \over {2M}}
-{{\left( {\hbar p} \right)^2} \over {2\mu }}}}dp}}} \right.
\\
\times \left[ {
     c_2
     \Delta _{kl}\left( {E-{{\left(\hbar q\right)^2}
     \over {2M }}-\hbar c\omega _\pi \left( Q \right)} \right)
     +c_2
     \Delta _{kl}\left( {E-{{\left( {\hbar q'} \right)^2} \over {2M}}
     -\hbar c\omega _\pi \left( Q \right)} \right)}
     +2\eta^{-1}_k\delta_{kl}
     \right]
\\
\left. {\times
     \int {{{\chi _l\left( {p'} \right)\chi _j\left( {p'} \right)}
     \over {E-{{\left( {\hbar q'} \right)^2} \over {2M}}-
     {{\left( {\hbar p'} \right)^2} \over {2\mu }}}}dp'}}\right\}
\, .
\end{multline}
This fully defines the driving term of the rank-N symmetrized
Lovelace-type equation.
\section{Bound-state equation}
\label{sec:homo}
In the previous sections we have provided the
AGS formulation for three-body scattering in one dimension.
However, we will now restrict the discussion to the corresponding
bound-state problem. We will apply this formalism
to the calculation of the binding energies, as a first test
for analyzing the effects of the irreducible pionic corrections
in the 1-D three-body system. As is known, the Faddeev or AGS method
formulates the scattering problem in terms of an inhomogeneous
integral equation which can be reduced, after iteration, to one of
Fredholm type. Such equations are known~\cite{Kowalski} to have no
solution in case the related homogeneous equation admits non-trivial
solutions, which are associated with bound states.

A further step, introduced for the convenience of numerical solution of the
homogeneous AGS-Lovelace Eq.~(\ref{eq:rankN}), is to express it in the
form of a sturmian eigenvalue problem~\cite{Sit91}. Therefore, in order
to find the three-body bound state energy, we solve the generalized eigenvalue
equation~\cite{PADC93}
\begin{equation}
\sum\limits_j {\int {{\cal U}_{ij}\left( {q,q',E} \right)\psi_{jk}
\left( {q'} \right)dq'}}=\eta _k\left( E \right)\sum\limits_j
{\int {{\cal Z}_{ij}
\left( {q,q',E} \right)\psi_{jk}\left( {q'} \right)dq'}}
     \label{eq:TGEV} \, .
\end{equation}
In this equation
\begin{equation}
{\cal U}_{ij}\left( {q,q',E} \right)=\sum\limits_{kl} {\int
{{\cal Z}_{ik}\left( {q,q'',E}
\right) \Delta_{kl}\left( q'';E\right){\cal Z}_{lj}
\left( {q'',q',E} \right)dq''}} \, ,
\end{equation}
and the remaining quantities are the same as those defined
in Sect.~\ref{sec:corr}. The eigenvalues
$\eta_k\left( E\right)$ are energy-dependent, because of the dependence
on the energy of ${\cal Z}$ and $\Delta$. The energy is varied, and the
bound-state energies of the three-body system are found by the condition that
the eigenvalue be
equal to one at $E=E_B$. Since the binding energy is negative,
all integrals (in Sect.~\ref{sec:corr}) involving the Green's function
are nonsingular, while for energies above the scattering
threshold, whenever such integrals become singular, they have to be
regularized with suitable subtraction techniques.
\section{Results of the Test Model}
\label{sec:num}
We have constructed three different two-body potentials that
mimic the nuclear
force on the line.
The general structure of these potentials is given by
\begin{equation}
   V\left( x\right)
  =V_{A}e^{-\beta \left|x\right|}+V_{R}e^{-n\beta \left|x\right|}
  \label{eq:MT-pot} \, ,
\end{equation}
with the free parameters $V_{A},V_{R}$.
In Table \ref{tbl:pots3} we report the three different sets of parameters.
The range parameter $\beta$ of the
attractive part is the inverse of the Compton wavelength
of the pion
\begin{equation}
\beta ={\frac{m_\pi c}{ \hbar }} \, ,
\label{eq:pion}
\end{equation}
which we consider fixed. The range parameter
of the repulsive part is variable and has been parametrized
in terms of multiples of $\beta$, denoted as $n$.
That is, $n$ denotes the ratio between the Compton wavelength
of the ``scalar pion'', the boson that mediates the attraction component,
and that of the repulsive short-range component. For the purpose
of the upcoming discussion it is unimportant whether the short-range part
is mediated by heavy-boson exchange, or by a generic cut-off mechanism
which suppresses the contributions coming from momenta higher than
the cut-off parameter. Three values for $n$ have been
selected, 5, 10, and 20,
corresponding to an increasing ``stiffness'' of the short-range repulsion.

The remaining parameters were fitted
to reproduce a two-body bound state at (approximately) the deuteron energy
and a two-body low-energy scattering parameter at about $24$~fm,
which would then correspond to the 1-D equivalent of the NN
scattering length.
\begin{table}[htb]
\beforetab
\begin{tabular}{ccccccc}
\firsthline
   & $n      $ & $V_{A}$ [MeV] & $V_{R}$ [MeV]
   &$ \hat \alpha_0 $  [fm]
   &$ \hat \alpha_1 $   [fm]  & $E_{B}$ [MeV]\\
\midhline
      $V_{5}$& 5 & -24.51 &   65.717  & 24.46 & -14.22 & -2.225 \\
\midhline
      $V_{10}$& 10 & -23.71  & 129.017 & 24.43 & -14.18 & -2.225 \\
\midhline
      $V_{20}$& 20 & -23.49  & 262.321 & 24.40 & -14.17 &  -2.225  \\
\lasthline
     \end{tabular}
\aftertab
\captionaftertab[]{Strength and range parameters of the
one dimensional potentials approximating nuclear physics and the
imaginary parts of the low energy expansion parameters \label{tbl:pots3}}
\end{table}
The binding energies reported in Table~\ref{tbl:pots3} were calculated
using the coordinate-space method described in
Appendix~C.
The low-energy scattering parameters $\hat \alpha_0,\hat \alpha_1$,
introduced in
Appendix~A, were calculated numerically
with the same method.

These potentials have a simple representation
also in momentum-space
\begin{equation}
V\left(k,k'\right)=\frac{\beta}{\pi}
\left(
\frac{V_A}{\left(k-k'\right)^2+\beta ^2}
+\frac{nV_B}{\left(k-k'\right)^2+\left(n\beta\right)^2}
\right)
\end{equation}
and it has been checked that the numerical calculations done
in momentum space
are consistent with the results obtained with
the coordinate-space method.

The bound-state energies of the three-body system are found by the
condition  that the generalized eigenvalue in Eq.~(\ref{eq:TGEV})
has to be equal to one. We
investigated the  effect of the correction terms for the three
potentials $V_5$, $V_{10}$,  and  $V_{20}$. Note that the coupling
constant of the force-mediating boson is implicitly defined
by the two-body potential, since the boson-exchange potential
is given by the exponential
\begin{equation}
{f^2\over\beta} e^{-\beta|x|}.
\end{equation}
This is the analogue of the Yukawa potential in one dimension.
Hence the correction terms discussed in section~\ref{sec:corr}
have to be evaluated with the condition $f^2=\beta V_A$.
\begin{figure}[hbt]
\begin{center}
{
\includegraphics[width=6cm]{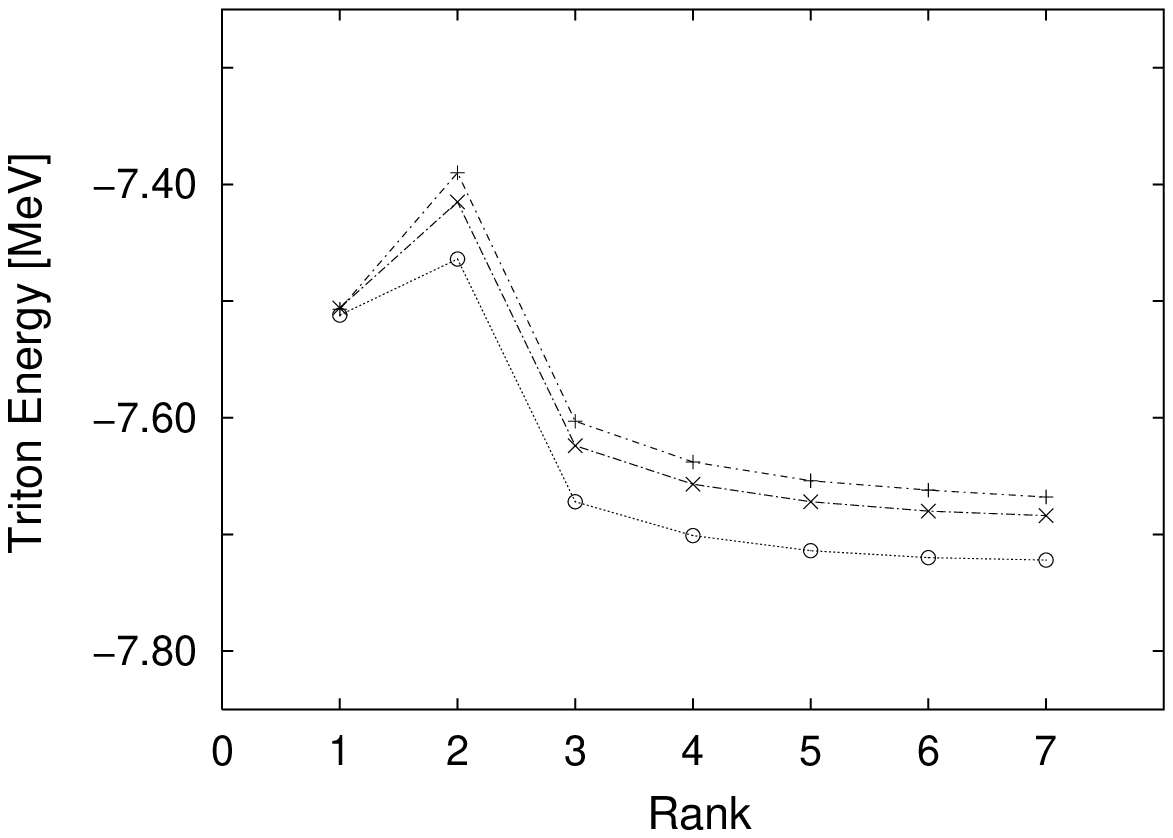}
\includegraphics[width=6cm]{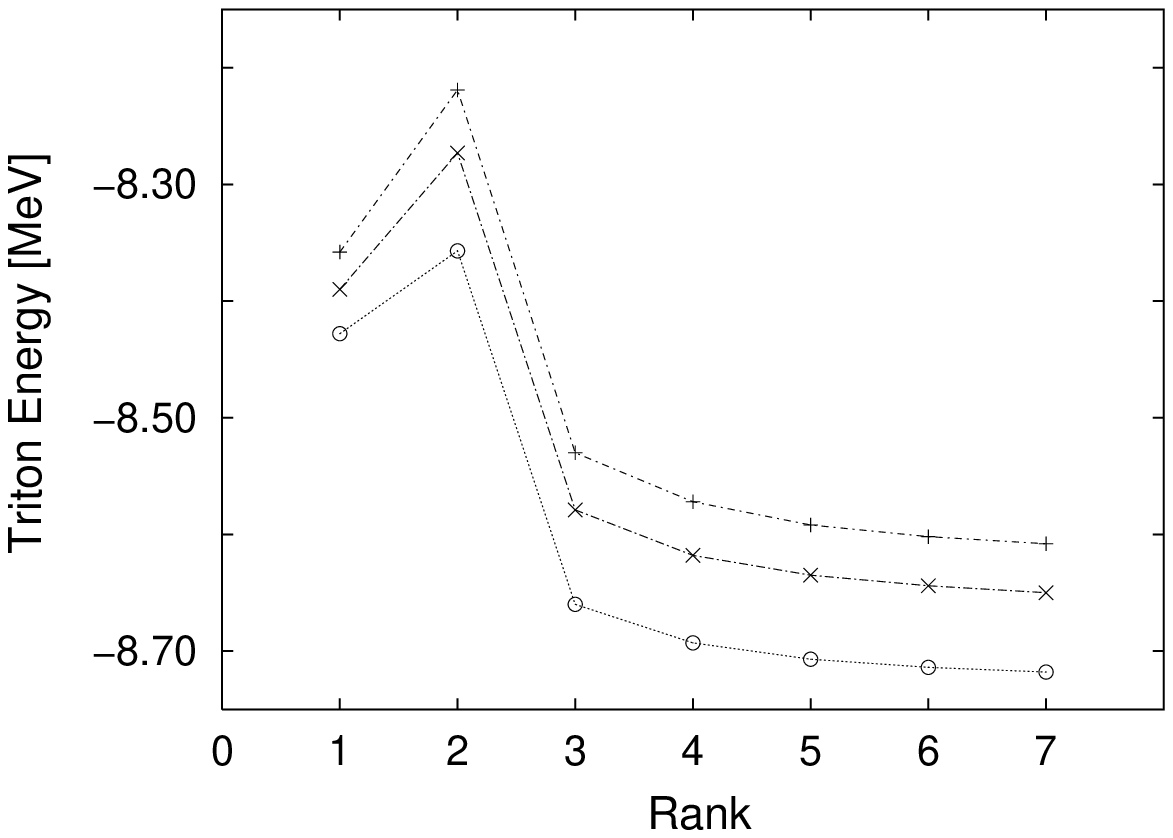}
}
\end{center}
\caption{Convergence behavior of the binding energy for the three potentials
in respect to the rank of the UPE. The left graph shows the calculation with
no correction terms and the right one with both correction terms
(diagonal and off-diagonal) included. The $\circ$ denote the results
for potential $V_5$, the $\times$ for potential $V_{10}$ and the $+$ for
potential $V_{20}$. }
\label{fig:conv}
\end{figure}

In Fig.~\ref{fig:conv} we
show the convergence behaviour of the three-body binding energy in
respect to the rank of the separable expansion for the three
potentials. The left graph shows  the convergence of the calculation
without any correction terms and the right  one with both corrections
(diagonal and off-diagonal). The two effective parameters
$c_1$ and $c_2$ have been set to $-15$ and $1$, respectively.
It is clearly seen that both calculations show a convergence in respect to
the rank of the UPE.
\begin{table}[hbt]
\beforetab
\begin{tabular}{ccccc}
\firsthline
          & 2NF & 2NF+OPE3 & 2NF+TPE3 & 2NF+OPE3+TPE3 \\
\midhline
   $V_5$                & -7.722 & -7.708 & -8.733 & -8.718 \\
\midhline
  $V_{10}$              & -7.684 & -7.662 & -8.670 & -8.650 \\
\midhline
  $V_{20}$              & -7.668 & -7.642 & -8.631 & -8.608 \\
\lasthline
     \end{tabular}
\aftertab
\captionaftertab[]{ `Triton' energies in MeV for the
three potentials, including
different 3NF corrections. The results are given for a rank 7 UPE
and with an effective parameter $c_2=1$ \label{tbl:Econv}}
\end{table}
In Table~\ref{tbl:Econv} we show the
numerical results for rank 7 calculations with
different combinations of correction terms.
It is worth noting that, once the local two-body potential has been
constructed to mimic the properties of the three-dimensional 2N system
(2N binding energy and low energy parameters), the three-body binding
energy turns out remarkably close to the values expected for the 3N system,
in spite of the fact that we have considered a rather different system,
namely three spinless bosons defined in one dimension.

It has also to be observed that the off-diagonal pionic contributions,
originated by the TPE3 diagram of Fig.~\ref{fig:FeynAGS}
can indeed be modelled
to correct a hypothetical under-binding of the three-body system, since
these terms have the potential to provide an additional  contribution to
the binding energy which can be sized around 0.9 MeV. This also
presents interesting analogies with the 3-D case of realistic 3N system.
We stress, however, that details can change, also significantly, depending
on the particular model considered for the off-shell extrapolation
of the $\pi$N amplitude.

The diagonal correction terms related to the OPE3 diagram
(Fig.~\ref{fig:Feyndiag})
produce changes in the three-body binding energies which are significantly
smaller than the off-diagonal ones. This also presents interesting analogies
with the realistic 3N case~\cite{CS01a}, where it has been already found that
the triton binding energy is not strongly affected by the diagonal
correction terms. Within this 1-D model one could conclude that
they could be neglected in first instance, if the effective
parameter $c_2$ is set to one.

Of course, this should not lead to the assumption that the OPE3
terms can also be ignored in calculations of other observables,
as it happens for the case of the vector analyzing powers,
for realistic nucleon-deuteron scattering at low energy.
It was shown in refs.~\cite{CS01a,CSH02} that the effective
parameter $c_2$ can be fitted to the peak of the $A_y$-data.
The value of the parameter in these fits depends sensitively on the
particular NN-potential used in the calculations, but in every case
the resulting curves reproduce the existing data base quite well.
Finally, it is worth mentioning that the two correction terms
produce additive effects to the three-body binding.

\begin{figure}[hbt]
\centerline
{
\includegraphics[width=6cm]{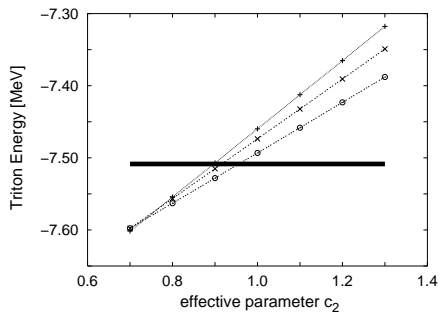}
}
\caption{Dependence of the binding energy on the parameter
$c_2$ in a rank 1 calculation for all three potentials including
the diagonal correction terms.
  The $\circ$ denote the results
for potential $V_5$, the $\times$ for potential $V_{10}$ and the $+$ for
potential $V_{20}$.
The thick line denotes the triton energy without any correction terms
for all three potentials. }
\label{fig:para}
\end{figure}
In Fig.~\ref{fig:para} we show how varying the parameter $c_2$  affects
the binding energy in the one-dimensional three-body model. The range of
variation for $c_2$ shown in the figure is consistent with the values
for $c_2$ found in refs.~\cite{CS01a,CSH02}. As mentioned earlier, one
possible reason for using $c_2$ as an adjustable parameter is the fact
that the 2N t-matrix entering in the OPE3 diagram has to be
calculated off the energy shell, in a region which is difficult to
constrain by experimental data. In addition, the figure reveals how the
variations with $c_2$ are dependent also on the selection of the
two-body potential.
At a parameter value of $0.7$ the diagonal correction terms introduce
an effect that is approximately $10\%$ of the effect introduced by the
off-diagonal correction terms.
At a value of $1.3$ the effect is in the opposite direction and
the magnitude of the effect is strongly dependent on the particular
potential, as seen by the different slopes. We have chosen to show the
rank $1$ calculation for different potentials, because the binding
energies without correction terms are practically equal for all three
potentials in this case. However, the same behavior can also be observed
in the higher-rank calculations. We conclude that, a priori, one should
not neglect the diagonal terms in this model calculation.

\section{Summary and Conclusion}
\label{sec:sum}
The scope of this study was originally the development of a strongly
simplified toy model in which to test a three-nucleon theory with residual
pionic corrections, namely with contributions that could not be
described via conventional two-nucleon potentials. The general
theoretical framework adopted as starting point has been the coupled
equations of ref.~\cite{Can98} for the $\pi$NNN-NNN system, further
reduced in ref.~\cite{CMS01} to an approximate but more practical
three-body dynamical equation with irreducible pionic corrections.

For this scope we formulated the quantum mechanical three-body problem
in one-dimension, on the $(-\infty,+\infty)$ line, and considered the
addition of two different 3NF-type diagrams generated by the  residual
pion dynamics. In the specialized literature, attention has been focused
--~also recently~-- to these types of diagrams in realistic 3N studies for
both bound and scattering regimes.

In this work, we concentrated our numerical study on the binding
energies of the three-body system.  We described the gross features of
the 3-body bindings, observing also interesting analogies with the more
realistic 3N-bound states, despite the rather severe simplifications
introduced by the one-dimensional model.

In particular, using simple 1-D potentials that could reproduce the
deuteron energy, we obtained  a 3-body binding of about 7.5 MeV.
Addition of residual pionic diagrams of the ``TPE3'' type
(Fig.~\ref{fig:FeynTM}) carries an  additional binding  of about
0.9 MeV.  However, significant model dependence on the structure of the
``$\pi$N'' amplitude (related to these diagrams) can and have to be
expected. The addition of the second type of diagrams, denoted
``OPE3'' (see Fig.~\ref{fig:Feyndiag}),  produces an additional effect on
the binding which is either negligible or moderate, depending on the
specifics of the selected two-body potential, and on the value of an
effective parameter that governs such diagrams. It must be observed that
one cannot conclude that the OPE3 terms are less important, since
they could quite significantly affect other three-body observables. This
indeed turned out to be the case for the nucleon-deuteron $A_y$
observable in realistic studies~\cite{CS01a,CSH02}.

However, we realized also that the 1-D approach  we
developed for the scopes discussed above went somehow beyond the
initial intentions of the authors. In fact it turned out, to the best of
our knowledge, that the three-body quantum scattering theory using the
Faddeev-AGS-Lovelace formulation in terms of coupled integral equations
has never been discussed before for one-dimensional systems,
while the general formulation for such systems could attract interest in
various fields dealing with one-dimensional systems.  For this reason,
we have devoted Sect.~\ref{sec:AGS} to the general discussion of the
quantum-mechanical 3-body scattering problem in one dimension. Both cases
involving distinguishable or identical particles have been formulated
in terms of the integral-equation approach.

As possible applications and future extensions of the 1-D model
considered herein, one could consider also other cases of three-body
models where there is the need of an explicit coupling to the meson
degree of freedom. This is the case, for example, of the constituent
quark model based on Goldstone-boson exchanges~\cite{GR96}, which provided
a rather realistic description of the excitation spectra of all light and
strange baryons~\cite{GPVW98}. One might think to improve the
description of the excitation spectra by including explicit mesonic
degrees of freedom into the resonance states, and in this framework to study
the $\pi$ and $\eta$ decay modes of N and $\Delta$ resonances.
Another possible application is in $\eta$-3N scattering~\cite{FA01}
and associated reactions~\cite{SBRSS01}. These processes are
dynamically coupled with the absorption channel, and with the
$\pi$-channel itself, and therefore the dynamics is highly complicated.
In both these examples, it would be rather useful to develop similar 1-D
models as theoretical laboratory where to test methods, approximations,
and dynamical assumptions in a simplified framework.

\begin{acknowledge}
  We acknowledge support from the Italian MURST-PRIN Project ``Fisica
  Teorica del Nucleo e dei Sistemi a Pi\`u Corpi''. T.M. acknowledges
  support from the University of Manitoba.
J.P.S. acknowledges support from the
  Natural Science and Engineering Research Council of Canada. The
  authors also would like to thank the Institutions of INFN (Padova),
  University of Manitoba, and Universit\`a di Padova for hospitality
  during reciprocal visits.
\end{acknowledge}
\appendix
\section{Partial Waves and Low Energy Behaviour in One-Dimensional Scattering}
\label{sec:LowE}
Here we review some general results obtained in 1-D scattering theory.
We consider a particle of mass $m$ with no internal degrees of freedom
subject to a potential $V\left( x\right)$ in one dimension. The potential is
assumed to be real and approaching zero sufficiently fast for
$x \rightarrow \pm \infty$. Furthermore, we restrict our investigation
to symmetric potentials $V\left( x \right)=V\left( -x \right)$. The
Schr\"odinger equation for this problem is
\begin{equation}
-\frac{d^2\Psi_k \left( x\right)}{dx^2}
+U\left( x\right)\Psi_k \left( x\right)
=k^2\Psi_k \left( x\right)
\, ,
\label{eq:Schroe}
\end{equation}
where $k=\left( 2m E\right)^\frac{1}{2}/ \hbar$ is the wave number,
$E$ is the energy of the particle and
$U\left( x\right)=\left(2m/ \hbar^2\right)V\left( x\right)$.
In this section we examine the
transmission-reflection problem for positive energies.
This problem has two independent solutions with
incident waves from the left and right respectively. The asymptotic form
for the wave function incident from the left is given by
\begin{equation}
\Psi_k^L\left( x\right)\rightarrow
\begin{cases}
e^{\imath kx}+R_L \left( E\right)e^{-\imath kx},
& \text{for $x \rightarrow -\infty$} \\
T_L \left( E\right)e^{\imath kx},
& \text{for $x \rightarrow +\infty$}
\, .
\end{cases}
\label{eq:Schroe-sol}
\end{equation}
The transmission and reflection coefficients $T_L,R_L$ lead to the
transmission and reflection probabilities, which have to satisfy
\begin{equation}
\left|T_L  \left( E\right)\right|^2+\left|R_L  \left( E\right)\right|^2=1
\end{equation}
due to conservation of probability. The asymptotic form for the wave
function incident from the right is defined similarly replacing $x$
with $-x$.\\
The S-matrix is given according to the definition~\cite{NR96,SB94}
\begin{equation}
S \left( E\right)=
\begin{pmatrix}
T_L  \left( E\right)& R_R  \left( E\right)\\
R_L  \left( E\right)& T_R  \left( E\right)\\
\end{pmatrix}
\, ,
\label{eq:Smatrix}
\end{equation}
which reduces to the identity matrix when the potential goes to zero.
In general, the matrix elements are complex and hence the matrix is defined
by eight real parameters. Unitarity of the S-matrix leads
to four constraints on the matrix elements and for real and symmetric
potentials two more constraints are observed. This leads to the identity
of the transmission and reflection coefficients for the incident waves
from the left and right
\begin{equation}
T \left( E\right)=T_L \left( E\right)=T_R \left( E\right)
\end{equation}
\begin{equation}
R \left( E\right)=R_L \left( E\right)=R_R \left( E\right)
\, ,
\end{equation}
which represent the only two independent parameters of the S-matrix
\begin{equation}
S \left( E\right)=
\begin{pmatrix}
T  \left( E\right)& R  \left( E\right)\\
R  \left( E\right)& T  \left( E\right)\\
\end{pmatrix}
\, .
\end{equation}
In the one-dimensional scattering problem there are two partial waves, one
with even parity and one with odd parity. The $\cal S$-matrix
in respect to these partial waves for real and symmetric potentials
is found by a diagonalization procedure
\begin{equation}
{\cal S} \left( E\right) =
\begin{pmatrix}
T \left( E\right)+R \left( E\right) & 0   \\
0   & T \left( E\right)-R  \left( E\right)\\
\end{pmatrix}
\, .
\end{equation}
Furthermore, due to the unitarity of the S-matrix, it can also be given
in the partial wave representation
\begin{equation}
{\cal S}  \left( E\right)=
\begin{pmatrix}
e^{2\imath\delta_0  \left( E\right)} & 0    \\
0 & e ^{2\imath\delta_1  \left( E\right)} \\
\end{pmatrix}
\, ,
\end{equation}
where $\delta_{0,1}\left( E\right)$ are real and define the phase shifts
of the scattering
process~\cite{Ebe65,For76}. In Fig.~\ref{fig:Phase} we show the odd
and even phase shifts calculated with the procedure described
in
Appendix~C.
\begin{figure}[hpt]
\centerline
{
\includegraphics[width=6cm]{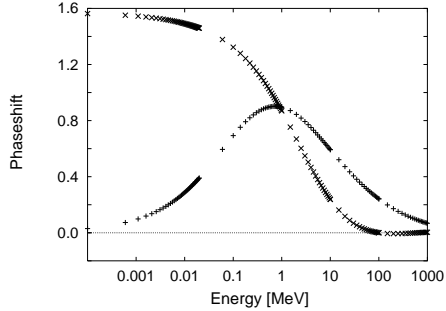}
}
\caption{Phase shifts for the odd and even wave function with potential
$V_{20}$, $+$ denotes the odd phase shift and $\times$ the even phase shift.}
\label{fig:Phase}
\end{figure}
They show the behaviour expected from the
one-dimensional Levinson theorem~\cite{SB94,BGK85}, namely the
odd phase goes to zero at threshold and the even to $\frac{\pi}{2}$,
because of the even bound state. For completeness, we note that
the transmission and
reflection coefficients can also be expressed in respect to the phase shifts
in the following way
\begin{eqnarray}
T \left( E\right) &=&  \frac{1}{2}\left(
e^{2\imath\delta_0\left( E\right)}+e ^{2\imath\delta_1\left( E\right)}
\right)
= \cos \left({\delta_0\left( E\right) -\delta_1\left( E\right) }\right)
e^{\imath\left(\delta_0\left( E\right) +\delta_1\left( E\right) \right)}
\\
R  \left( E\right) &=&  \frac{1}{2}\left(
e^{2\imath\delta_0\left( E\right)}-e^{2\imath\delta_1\left( E\right)}
\right)
=\sin \left({\delta_0\left( E\right) -\delta_1\left( E\right) }\right)
e^{\imath\left(\delta_0\left( E\right) +\delta_1\left( E\right)
+\frac{\pi}{2} \right)}
\, .
\end{eqnarray}
Finally, according to
Galindo and Pascual~\cite{GP90}, for potentials that go to
zero sufficiently fast, when $x$ goes to plus
or minus infinity, the following generic behavior is observed
\begin{eqnarray}
T\left( E \right) & \approx & \alpha_0 \sqrt{\frac{2mE}{\hbar^2}},
\qquad \qquad \: \;  \sqrt{\frac{2mE}{\hbar^2}}\rightarrow 0
\label{eq:T-low-e}\\
R\left( E \right) & \approx & -1 +\alpha_1  \sqrt{\frac{2mE}{\hbar^2}},
\qquad  \sqrt{\frac{2mE}{\hbar^2}}\rightarrow 0
\label{eq:R-low-e}
\, ,
\end{eqnarray}
where $\alpha_0,\alpha_1$ are generally complex constants not equal to zero.
Therefore, the S-matrix in the low-energy limit tends to a total reflection.
It should be noted that, for the specific type of potentials we use,
it turns  out the constants $\alpha_0, \alpha_1$ are purely imaginary
numbers.

\begin{figure}[hbt]
\centerline
{
\includegraphics[width=6cm]{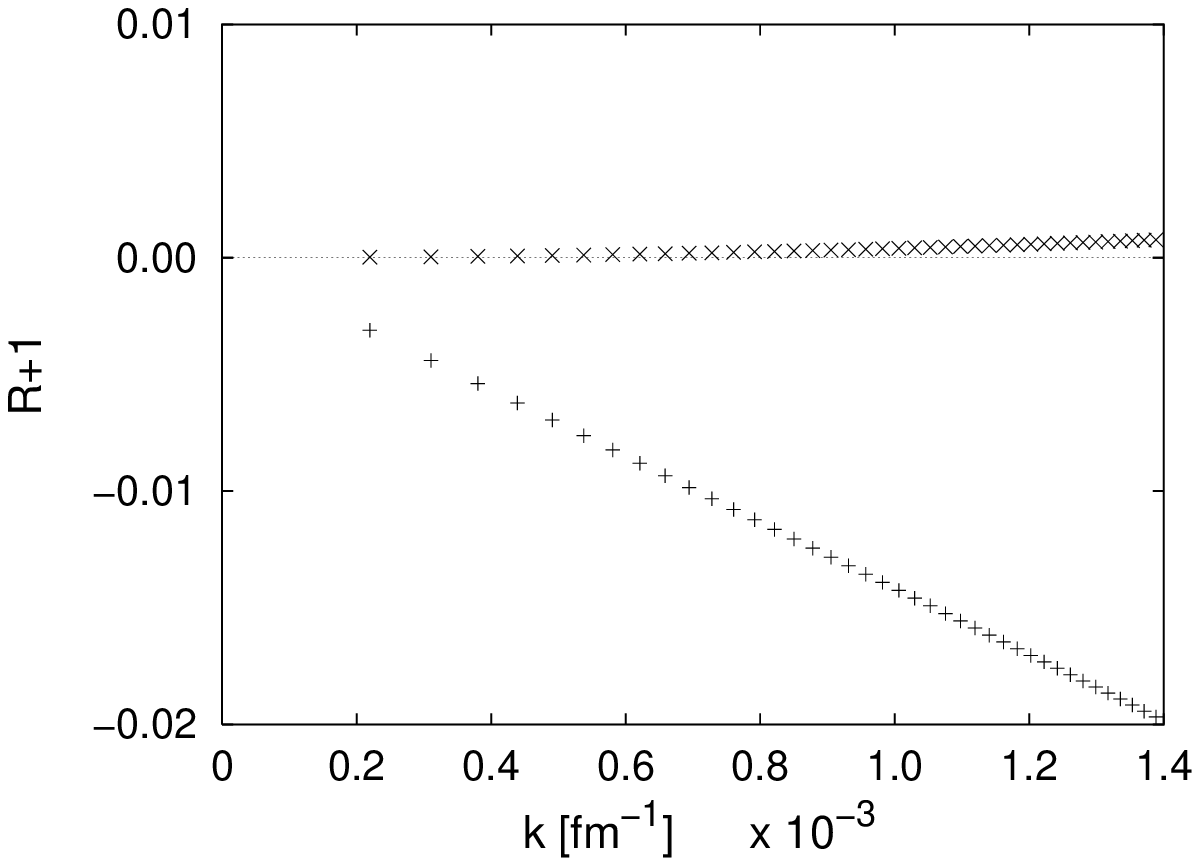}
\includegraphics[width=6cm]{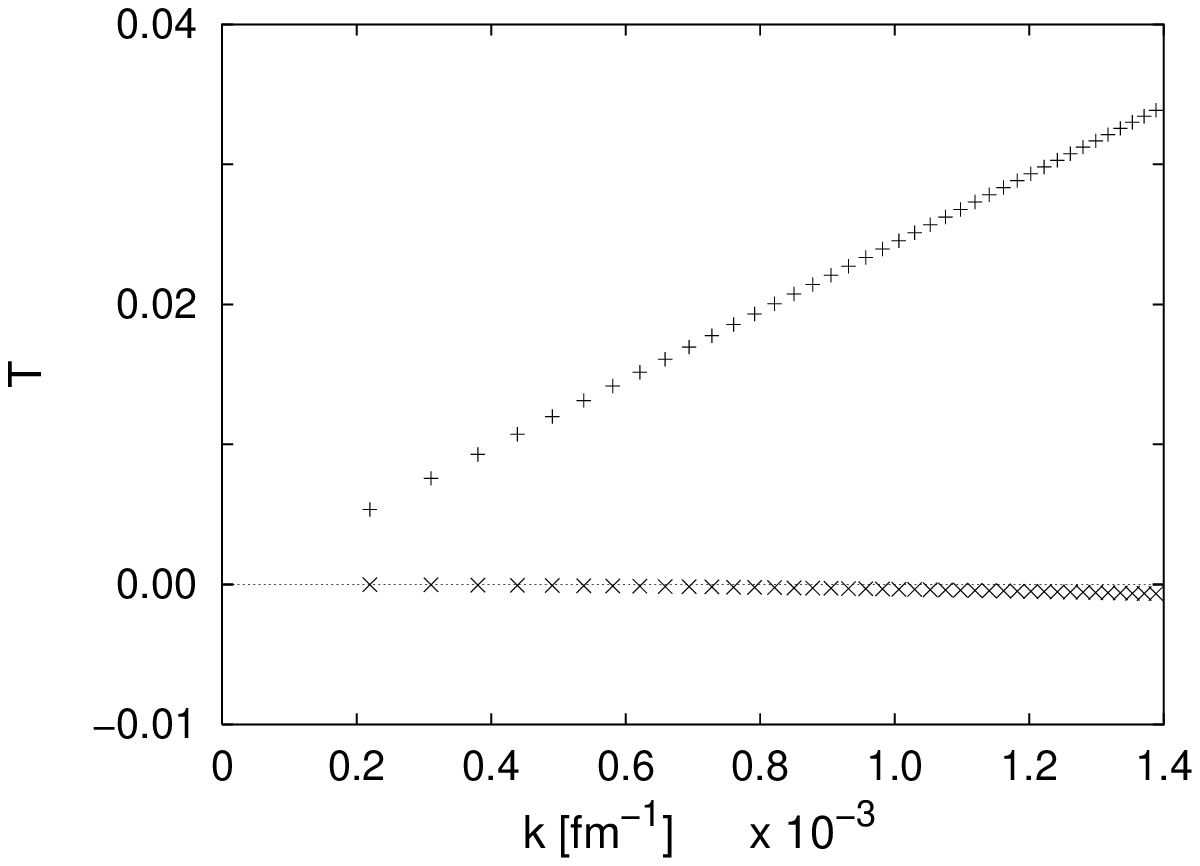}
}
\caption{Real and imaginary parts of the coefficients  $R+1$ and $T$.
The $+$ denote the imaginary parts of the
coefficients, the $\times$ denote the real parts of the
coefficients.
}
\label{fig:transcoeff}
\end{figure}

The transmission and reflection coefficients can be calculated by
resorting to
the Green's function formalism~\cite{Mel01,BLA01}. The Schr\"odinger
Eq.~(\ref{eq:Schroe}) can be written in integral form, which leads
to the Lippmann-Schwinger equation
\begin{equation}
\Psi_k\left( x\right)=\Phi_k\left( x\right)+\int_{-\infty}^{+\infty}
{
G_0\left(x,x'\right)U\left( x'\right)\Psi_k\left( x'\right)dx'
}
\, ,
\end{equation}
where $\Phi_k\left( x\right)$ is a solution of the free Schr\"odinger
equation and the Green's function $G_0\left(x,x'\right)$ is defined
incorporating the boundary conditions of
the equivalent Schr\"odinger problem.
For 1-D systems the Green's function with outgoing boundary conditions has
the following form
\begin{equation}
G_0\left(x,x'\right)=-\frac{\imath}{2k}e^{i k\left|x-x'\right|}
\, .
\end{equation}
The general solution of the free Schr\"odinger Eq.~(\ref{eq:Schroe})
is given by
\begin{equation}
\Phi_k\left( x\right)=Ae^{\imath kx}+Be^{-\imath kx}
\, ,
\end{equation}
where the coefficients $A$ and $B$ are arbitrary.
Inserting these expressions
in the Lippmann-Schwinger equation leads directly to the following equation
\begin{multline}
\Psi_k\left( x\right) = Ae^{\imath kx}+Be^{-\imath kx}\\
-\frac{\imath m}{\hbar^2 k}\left[\int_{-\infty}^{x}
{
e^{\imath k\left(x-x'\right)}V\left(x'\right)\Psi_k\left( x'\right)dx'
}
+\int_{x}^{+\infty}
{
e^{\imath k\left(x'-x\right)}V\left(x'\right)\Psi_k\left( x'\right)dx'
}
\right]
\label{eq:TR2spaceprelim}
\end{multline}
and $A,B$ can now be determined according to the various boundary conditions.
One choice is $A=1,B=0$ and corresponds to the following equation
\begin{eqnarray}
\Psi_k^{L}\left( x\right)& = & e^{\imath kx}-\frac{\imath  m}{\hbar^2 k}\left(
\int_{-\infty}^{+\infty}e^{\imath kx'}V\left( x'\right)
\Psi_k^{L}\left( x'\right)dx'\right)
e^{-\imath kx} \qquad x\rightarrow -\infty
\\
\Psi_k^{L}\left( x\right)& = & \left[ 1-\frac{\imath m}{\hbar^2 k}\left(
\int_{-\infty}^{+\infty}e^{-\imath kx'}V\left( x'\right)
\Psi_k^{L}\left( x'\right)dx'\right)\right]
e^{\imath kx} \qquad x\rightarrow +\infty
\, ,
\end{eqnarray}
which is equivalent to the previously defined solution in
Eq.~(\ref{eq:Schroe-sol}). The transmission and reflection
coefficients for scattering with
an incoming wave from the left are then given by the expressions
\begin{eqnarray}
T_L  \left( E\right)& = & 1+\frac{\imath}{2k}f_+^{kL}
\label{eq:Tfs1}\\
R_L  \left( E\right)& = & \frac{\imath}{2k}f_-^{kL}
\, ,
\label{eq:Tfs}
\end{eqnarray}
where
\begin{equation}
f_\pm^{kL} = -\frac{2m}{\hbar ^2 }
\int_{-\infty}^{+\infty}{e^{\mp \imath x'k}
V\left( x'\right)\Psi_k^L\left( x'\right)dx'}
\end{equation}
is the scattering amplitude for an incoming wave from the left.
In a similar way it is possible to describe
the scattering system with an incoming wave from the right by choosing the
coefficients $A=0,B=1$. However, we already argued that for the type of
potentials we use in this paper the left- and right- coefficients are equal.
Therefore, we drop the indices $L,R$ in the remainder of this section and
denote the scattering amplitude in both directions by $f_\pm^{k}$.
Observing that in the one-dimensional system, for a given energy $E$,
there exist exactly two possible momenta
\begin{equation}
k_\pm=\pm \frac{\sqrt{2mE}}{\hbar}
\end{equation}
we define the on-shell t-matrices by the expression
\begin{eqnarray}
t\left(k_\pm,k;E\right) &= &
\left<k_\pm\right|t\left( E\right)\left|\Phi_k\right>=
\left<\Phi_{\pm k}\right|V\left|\Psi_k\right>
\nonumber \\
&=&\int_{-\infty}^{+\infty}
{
e^{\pm \imath kx'}V\left( x'\right)\Psi_k\left( x'\right)dx'
}
=-\frac{\hbar^2}{2m}f^k_\pm \, .
\label{eq:tmat2space}
\end{eqnarray}
For the off-the-energy-shell situation the t-matrix is introduced according
to
\begin{equation}
t\left( E\right)\left|\Phi_q\right>=V\left|\Psi_q\right>
\end{equation}
and it can be calculated by the Lippmann-Schwinger equation
\begin{equation}
t\left( E\right)=V+VG_0t\left( E\right)
\label{eq:LS-append}
\end{equation}
which reads in momentum-space
\begin{equation}
t\left(q,q';E\right)=V\left(q,q'\right)+\int_{-\infty}^{+\infty}
{
V\left(q,q''\right)G_0\left(q'',E\right)t\left(q'',q';E\right)dq''
}
\end{equation}
where
\begin{equation}
G_0\left(q,E\right)=\left( E-\frac{\hbar^2q^2}{2M}\right)^{-1}
\end{equation}
Finally, combining
Eqs.~(\ref{eq:Smatrix},\ref{eq:Tfs1},\ref{eq:Tfs},\ref{eq:tmat2space})
it is possible to express the S-matrix in respect to the on-shell t-matrix
\begin{equation}
S\left( E\right)=
\begin{pmatrix}
1 & 0\\
0 & 1
\end{pmatrix}
-\frac{\imath m}{\hbar^2 k}
\begin{pmatrix}
t\left(k,k;E\right)  & t\left(k,-k;E\right)\\
t\left(-k,k;E\right) & t\left(-k,-k;E\right)
\end{pmatrix}\, .
\label{eq:STconn2space}
\end{equation}
A diagonalization procedure lets us express this result in terms of
two ``partial wave'' t-matrices
\begin{equation}
{\cal S}\left( E\right)=
\begin{pmatrix}
1 & 0\\
0 & 1
\end{pmatrix}
-\frac{\imath m}{\hbar^2 k}
\begin{pmatrix}
t_0\left(E\right)  & 0\\
0 & t_1\left(E\right)
\end{pmatrix} \, ,
\end{equation}
where
\begin{eqnarray}
t_0\left(E\right)& = &t\left(k,k;E\right)+t\left(k,-k;E\right)\\
\nonumber
t_1\left(E\right)& = &t\left(k,k;E\right)-t\left(k,-k;E\right)
\end{eqnarray}
The low energy behavior of the t-matrix can be defined by a direct comparison
to the low energy behavior of the on-shell S-matrix, which yields
\begin{eqnarray}
t_0 \left( E\right) & \approx &
-\frac{\hbar^2}{i m}\sqrt{\frac{2mE}{\hbar^2}}
\left[\left(\alpha_0+\alpha_1\right)\sqrt{\frac{2mE}{\hbar^2}}-2\right]
\label{eq:low-e-behave}\\
t_1 \left( E\right)  & \approx &
-\frac{\hbar^2}{i m}\sqrt{\frac{2mE}{\hbar^2}}
\left(\alpha_0-\alpha_1\right)\sqrt{\frac{2mE}{\hbar^2}}
\, ,
\label{eq:low-e-behave1}
\end{eqnarray}
where $0,1$ denote the even and odd t-matrix respectively.
In Fig.~\ref{fig:transcoeff} we show the low energy behaviour of the
transmission and reflection coefficients explicitly.
The figures were
calculated using the procedure outlined in
Appendix~C.
Clearly, the imaginary parts of the
reflection and transmission coefficients show the linear behaviour,
while the real parts are equal to zero.
In the TPE3 a subtracted $\pi$N t-matrix enters and we approximate it
by the definition
\begin{equation}
t_F \left( q,q'\right)  =  -\frac{\hbar^2}{ m}\left[
\frac{\left(\hat\alpha_0+\hat\alpha_1\right)\left|q\right|\left|q'\right|
+\left(\hat\alpha_0-\hat\alpha_1\right)qq'}{2}\right]
\, .
\end{equation}
Here, the parameters $\hat \alpha_0,\hat \alpha_1$ denote the imaginary parts
of $\alpha_0,\alpha_1$,
which are given by the Eqs.~(\ref{eq:T-low-e},\ref{eq:R-low-e}).
Furthermore, to get this result from
Eqs.~(\ref{eq:low-e-behave},\ref{eq:low-e-behave1}),
it must be
observed that both odd and even terms must be added coherently, since
the meson and the nucleon are distinguishable.
Finally, we have neglected the additional contribution due to the
presence of the term $-1$ in Eq.~(\ref{eq:R-low-e}).
This contribution originates because in the low-energy limit
the 1-D scattering
event becomes a purely reflective process ($T\rightarrow 0$,
$R\rightarrow -1$). It is eliminated from Eq.~(\ref{low-exp})
because in the $\pi$N t-matrix there are contributions that need to
be subtracted, i.e. the forward propagating nucleon
diagram and its crossed counterpart. In the three-dimensional case, the
subtraction of such diagrams is discussed for example
in ref.~\cite{van94}. Finally, a phenomenological
constant $c_1$ has to be introduced to account for the different magnitudes
of the
low energy $\pi$N and NN t-matrices. For insertion
into the off-diagonal correction terms the expression is rewritten
in the appropriate units multiplying it by the constant
$\left(2m\right)/ \hbar^2$.
\section{The unitary pole expansion}
\label{sec:UPE}
In this appendix we describe the Unitary Pole Expansion (UPE), which
is the input for our three-body bound state problem.  The UPE is
closely related to the Weinberg series~\cite{Wei63} and is described
by Harms~\cite{Har67}. We derive the UPE in a slightly different, but
equivalent, way as discussed by Pisent et al.~\cite{PADC93}.
The homogeneous form of the Lippmann-Schwinger Equation~(\ref{eq:LS-append})
can be given by the generalized eigenvalue equation
\begin{equation}
VG_0\left( E \right)V\left| {\phi _n\left( E \right)} \right\rangle
=\eta _n\left( E \right)V\left| {\phi _n\left( E \right)} \right\rangle
\, .
\label{eq:GEV}
\end{equation}
The form factors of the separable expansion are defined by
\begin{equation}
\left| {\chi _n\left( E \right)} \right\rangle =
V\left| {\phi _n\left( E \right)} \right\rangle
\, .
\end{equation}
In our case we have only one bound state and we choose to use the
expansion at
the bound state energy $E=-B$. In this way we recover the Unitary Pole
Expansion (UPE). Furthermore, it is easily seen that the first
eigenvalue $\eta _1\left( E \right)$ is equal to one in this case.
Therefore, the bound state $\left| {\phi _B} \right\rangle $ and the
first form factor in the expansion satisfy the relation
\begin{equation}
\left| {\chi _1} \right\rangle =V\left| {\phi _B} \right\rangle
\, .
\end{equation}
The bound state calculated with this definition was compared to the
one calculated with the Fox-Goodwin method and it was seen that they
indeed do agree with each other.
It is also known~\cite{PADC93} that the eigenvectors of
Eq.~(\ref{eq:GEV})
have the following orthogonality relation
\begin{equation}
\left\langle {\phi _n} \right|V\left| {\phi _m} \right\rangle =
-\eta _n\delta _{nm}
\, ,
     \label{eq:sturm:orth}
\end{equation}
where the $\left| {\phi _m} \right\rangle $ are the solutions of
Eq.~(\ref{eq:GEV}) at the binding energy $E=-B$.
This orthogonality relation
also specifies the normalization of the eigenfunctions, which is slightly
different from that introduced in Section \ref{sec:AGS}.
The two-body
potential $V$ is given in respect to the following separable expansion
\begin{equation}
V=-\sum\limits_i{\frac{\left|\chi_i\right>\left<\chi_i\right|}{\eta_i}}
     \label{eq:sep-pot}
\end{equation}
and the separable T matrix is given by
\begin{equation}
T=\sum\limits_{i,j}{\left|\chi_i\right\rangle \Delta _{ij}
\left\langle \chi_j \right|}
     \label{eq:sep-t}
\, ,
\end{equation}
where the matrix elements of $\Delta$ are defined by the expression
\begin{equation}
\left[ {\Delta ^{-1}} \right]_{ij}=-\eta _i\delta _{ij}-
\left\langle {\chi _i} \right|G_0\left| {\chi _j} \right\rangle
\, .
     \label{eq:tau1}
\end{equation}
In the case that only one term in this expansion is retained the
expansion reduces to the Unitary Pole Approximation (UPA) discussed by
Lovelace~\cite{Lov64} and Fuda~\cite{Fud68}
\begin{equation}
T_{UPA}=\left| {\chi _1} \right\rangle \Delta _{11}
\left\langle {\chi _1} \right|
\end{equation}
with
\begin{equation}
\Delta _{11}=-\left[ {1+\left\langle {\chi _1} \right|G_0
\left| {\chi _1} \right\rangle } \right]^{-1}
\, .
\end{equation}
It is easily seen that both the UPE and UPA have the right pole
structure, namely the separable T matrix has a pole at the bound state
energy.
\section{Numerical Studies of the Two-Body Problem in Coordinate Space}
\label{sec:1Dx}
We note that our one-dimensional system is
defined on the whole line and not the half line like the radial
Schr\"odinger equation of a three-dimensional problem.
In this Appendix we show how the system can
be solved numerically for any finite-range symmetric potential.\\
The problem is defined by the time-independent Schr\"odinger equation
\begin{equation}
H\psi \left( x \right)=E\psi \left( x \right)
\, ,
\end{equation}
which can be written as
\begin{equation}
u''\left( x \right)+w\left( x \right)u\left( x \right)=0
\, ,
\end{equation}
where
\begin{equation}
w\left( x \right)={{2m} \over \hbar^2 }\left( {E-V\left( x \right)} \right)
\, .
\end{equation}
The Fox-Goodwin method \cite{FG88} replaces the differential equation
by a finite expression
\begin{equation}
u\left( {x+h} \right)={\frac{2u\left( x \right)-u\left( {x-h} \right)
-\left( {{\frac{h^2} {12}}} \right)\left( {10w\left( x \right)u
\left( x \right)+w\left( {x-h} \right)u\left( {x-h} \right)} \right)}
{1+\left( {{\frac{h^2} {12}}} \right)w\left( {x+h} \right)}}
\, ,
\label{eq:Foxmain}
\end{equation}
which is exact to the order of $h^{4}$. If one knows the values of
the function $u$ at the positions $0$ and
$h$, then it is possible to calculate the value of the function at the
point $2h$. Repeating this procedure with the function at $h$ and $2h$
will give its value at $3h$ and so on. With this procedure the function
is defined on a grid with discretization interval $h$.
The first two values chosen in this procedure, namely
$u\left(0\right)$ and $u\left(h\right)$ depend on the boundary
conditions of the system. This procedure works well for the
radial equation, because one has a definite starting point at zero.

The generalization to the full line is simplified if one
takes advantage of the symmetric nature of the potential.
The symmetry implies that the eigenstates can be classified into
symmetric (even) or anti-symmetric (odd) states.
Therefore, it is possible to
restrict the calculation to the
positive part of the real line by imposing
the appropriate boundary
conditions at the origin for both odd and even states.

Odd states have to go through zero
at the origin.
Consequently, we define the boundary conditions at
the origin in the following way
\begin{equation}
u_1\left( 0 \right)= 0 \qquad u_1\left( h \right) = 1
\, .
\label{evenstart}
\end{equation}
Even states have to go through an
extremum at zero, but without crossing the origin.
Consequently, we define the symmetric boundary conditions
as follows ($h\rightarrow 0$)
\begin{equation}
u_0\left( 0 \right)=1 \qquad u_0\left( h \right)=1
\, .
\label{oddstart}
\end{equation}
\subsection{The Bound State Problem on the Real Line}
Bound-state solutions, in both odd and even cases, have to be
normalizable. In the positive direction, negative-energy solutions
follow the asymptotic behaviour
\begin{equation}
u\left( x \right)=ae^{-\kappa x}+be^{\kappa x}
\, ,
\label{as-bound}
\end{equation}
with
$\kappa={\sqrt{2m\left|E\right|}}/\hbar$
being the wave number for $E<0$.

The
normalization condition determines the bound-state
by selection of the energies corresponding to a zero of the
$b$ coefficient, namely a vanishing wavefunction for
$x\rightarrow +\infty$. Numerically, the bound states and
energies are found by starting the Fox-Goodwin algorithm
at $x=0$ with either the odd and even
Conditions~(\ref{oddstart},\ref{evenstart}),
and then evaluating the wavefunction for very large
$x$ values.  Generally, the function assumes very large
values, due to the increasing exponent
in Eq.~(\ref{as-bound}),
except for energy values corresponding to bound states,
where the wavefunction drops to zero at large distances
and becomes normalizable.
\subsection{The Scattering Problem on the Real Line}
The 1-D scattering problem is solved separately for antisymmetric
and symmetric states by starting at the origin with either
Condition~(\ref{oddstart}) or (\ref{evenstart}), and propagating the
solution in the asymptotic region $x\rightarrow +\infty$
with the Fox-Goodwin method.
In the odd case we know that the
solution in the asymptotic region
has to be of the form
\begin{equation}
\bar u_{1}\left( x \right)=a_{1}\sin \left( {kx+\delta_{1}} \right)
\label{asymtodd}
\end{equation}
with $k={\sqrt{2mE}}/\hbar$ being
the wave number and the bar denoting the asymptotic solution.

Equating the last two points of the numerical solution with this
general expression gives the following two
equations
\begin{eqnarray}
u_{1}\left( B \right)&= &a_{1}\left[ {\cos \left( {\delta _1} \right)
\sin \left( {kB} \right)+\sin \left( {\delta _1} \right)
\cos \left( {kB} \right)} \right]
\nonumber
\\
u_1\left( {B-h} \right) & =& a_1\left[ {\cos \left( {\delta _1}
\right)\sin \left( {k\left( {B-h} \right)} \right)+
\sin \left( {\delta _1} \right)\cos \left( {k\left( {B-h} \right)}
\right)} \right]
\, ,
\label{ODD-SYSTEM}
\end{eqnarray}
where $B$ and $B-h$ are the largest and second largest $x$-value at which
the wavefunction is calculated numerically. Solution of this system leads to
the following expression defining the phase shift
\begin{equation}
\tan \left( {\delta _1} \right)=
{\frac{\sin \left[ {k\left( {B-h} \right)} \right]
u_1\left( B \right)-\sin \left[ {kB} \right]
u_1\left( {B-h} \right)} {-\cos \left[ {k\left( {B-h} \right)} \right]
u_1\left( B \right)+\cos \left[ {kB} \right]
u_1\left( {B-h} \right)}}
\, .
\end{equation}
Substituting the phase shift back into~(\ref{ODD-SYSTEM})
gives also the odd amplitude $a_1$. \\

The even phase shift can be similarly determined
by the asymptotic behavior
\begin{equation}
\bar u_0\left( x \right)=a_0\cos \left( {kx+\delta_0} \right)
\, ,
\label{asymteven}
\end{equation}
which yields
\begin{equation}
\tan \left( {\delta _0} \right)={\frac{\cos \left[ {kB} \right]
u_0\left( {B-h} \right)-\cos \left[ {k\left( {B-h} \right)} \right]
u_0\left( B \right)} {\sin \left[ {kB} \right]
u_0\left( {B-h} \right)-\sin \left[ {k\left( {B-h} \right)} \right]
u_0\left( B \right)}}
\, .
\end{equation}
As for the antisymmetric case, also the symmetric amplitude $a_0$ can be
determined in the same way.

Generally, the asymptotic behavior of the one dimensional
scattering system for incoming waves from the left ($x\rightarrow -\infty$),
  is described by
\begin{eqnarray}
  \bar u\left( x\rightarrow -\infty \right)
& = & e^{\imath kx}+{ R}\cdot e^{-\imath kx}
\label{eq:-inf}\\
\bar u\left( x\rightarrow +\infty \right)
& = &{ T}\cdot e^{\imath kx}
\, ,
     \label{eq:+inf}
\end{eqnarray}
where $R$ and $T$ denote the reflection and
transmission coefficients. We define the scattering wavefunction $u(x)$
with the proper asymptotic behaviour
by a (suitably chosen) linear combination of odd and even solutions, namely
\begin{equation}
u\left( x \right)=\frac{e^{\imath\delta _0}}{a_0}u_0\left( x \right)
+\imath \frac{e^{\imath\delta _1}}{a_1}u_1\left( x \right)
  \, .
\end{equation}
Comparing the asymptotic behaviour for both antisymmetric and symmetric
Solutions~\ref{asymtodd}~\ref{asymteven} with the
Form~\ref{eq:-inf}~\ref{eq:+inf} one derives the two scattering coefficients
\begin{eqnarray}
T \left( E\right) &=&  \frac{1}{2}\left(
e^{2\imath\delta_0\left( E\right)}+
e^{2\imath\delta_1\left( E\right)}
\right)
\\
R  \left( E\right) &=&  \frac{1}{2}\left(
e^{2\imath\delta_0\left( E\right)}-
e^{2\imath\delta_1\left( E\right)}
\right)
\, ,
\end{eqnarray}
according to the expressions already mentioned in
Appendix~A.

\end{document}